\documentclass[aps,prb,reprint,amsmath,amssymb,superscriptaddress]{revtex4-2}
\usepackage{graphicx}
\usepackage{dcolumn}
\usepackage{bm,epsfig}
\usepackage{epstopdf}
\usepackage{xcolor}
\usepackage{float}
\usepackage{tabularx}
\usepackage{hyperref}

\begin{document}

\title{Pressure evolution of coplanar antiferromagnetism in heavy-fermion Ce$_{2}$CoAl$_{7}$Ge$_{4}$}

\author{M. O. Ajeesh} \email{ajeesh@iitpkd.ac.in} \affiliation{Los Alamos National Laboratory, Los Alamos, New Mexico 87545, USA} \affiliation{Department of Physics, Indian Institute of Technology Palakkad, Kerala 678623, India}
\author{A. O. Scheie} \email{scheie@lanl.gov} \affiliation{Los Alamos National Laboratory, Los Alamos, New Mexico 87545, USA}
\author{Yu Liu} \affiliation{Los Alamos National Laboratory, Los Alamos, New Mexico 87545, USA}
\author{L. Keller}  % lukas.keller@psi.ch
\affiliation{Laboratory for Neutron Scattering and Imaging, Paul Scherrer Institut, Villigen CH-5232, Switzerland}
\author{S.~M.~Thomas} \affiliation{Los Alamos National Laboratory, Los Alamos, New Mexico 87545, USA}

\author{P.~F.~S.~Rosa} \affiliation{Los Alamos National Laboratory, Los Alamos, New Mexico 87545, USA}

\author{E.~D.~Bauer} \affiliation{Los Alamos National Laboratory, Los Alamos, New Mexico 87545, USA}
\date{\today}

\date{\today}

\begin{abstract}

Ce$_{2}$$M$Al$_{7}$Ge$_{4}$ ($M=$ Co, Ir, Ni or Pd) are heavy-fermion materials and host a variety of ground states ranging from magnetism to non-Fermi liquid behavior. The Co, Ir, and Ni members of the series undergo magnetic ordering with decreasing transition temperatures. In contrast, the Pd compound does not magnetically order down to 0.4 K and shows non-Fermi liquid behavior, suggesting proximity to a magnetic quantum critical point. Among the series, Ce$_{2}$CoAl$_{7}$Ge$_{4}$ orders antiferromagnetically below $T_N=1.9$~K along with heavy-Fermion behavior below 15 K. We report the magnetic structure of the antiferromagnetic phase in Ce$_{2}$CoAl$_{7}$Ge$_{4}$ and the evolution of the magnetic transition under external pressure. Rietveld refinement of the neutron diffraction data suggests a coplanar antiferromagnetic structure with a wave vector $k = (1,1,1)$ and an ordered moment of $0.383 \pm 0.018 \> \mu_B$ in the antiferromagnetic phase. Electrical resistivity and AC calorimetry measurements under hydrostatic pressure reveal a suppression of the antiferromagnetic transition toward zero temperature around $p=1.1$ GPa. However, there is no evidence of non-Fermi liquid behavior associated with the suppression of magnetism by pressure, unlike the effect of transition-metal substitution.

\end{abstract}

\maketitle

\section{Introduction}
Ce-based heavy-fermion materials exhibit a variety of ground states due to the competition between the Kondo effect and the Ruderman–Kittel–Kasuya–Yosida (RKKY) interactions~\cite{Doniach77,Stewart84,Fizk88,Coleman07}. The competition between these interactions is delicately controlled by the strength of the hybridization between the $4f$ electrons and the conduction electrons, which can be tuned by nonthermal control parameters such as chemical substitution, external pressure, and magnetic field. Suppressing magnetic ordering in such materials to zero temperature may result in a quantum critical point (QCP), where the properties are governed by quantum fluctuations~\cite{Sachdev00,Gegenwart08, Stockert11}. Magnetic instabilities near the QCPs often give rise to unconventional phases and phenomena, including superconductivity and non-Fermi liquid behavior~\cite{Mathur98,Steglich14,Stewart01,Loehneysen07}. Identifying candidate materials and tuning their ground states is crucial for advancing our understanding of these unconventional phenomena.

The Ce$_{2}$$M$Al$_{7}$Ge$_{4}$ family of compounds, where $M =$ Co, Ir, Ni, or Pd, shows heavy-fermion behavior and a variety of ground states depending on the transition metal~\cite{Ghimire16}. These compounds crystallize in the non-centrosymmetric tetragonal structure with the space group $P\bar{4}2_1m$. Earlier studies have shown that chemical substitution at the transition metal site suppresses the magnetic transition temperature from 1.8 to 0.8 K as it goes from Co to Ni. Interestingly, no magnetic order was found in the Pd compound for temperatures down to 0.4 K. In addition, low-temperature heat capacity, electrical resistivity, and magnetic susceptibility data show signatures of non-Fermi liquid behavior in the Pd compound~\cite{Ghimire16}. These results suggest that the Ce$_{2}$$M$Al$_{7}$Ge$_{4}$ series is close to a magnetic quantum critical point and therefore serves as an ideal candidate for investigating pressure-induced quantum criticality. 

Among the 2174 family, Ce$_{2}$CoAl$_{7}$Ge$_{4}$ is reported to order antiferromagnetically around $T_N=1.8$ K, while the Ir and Ni compounds order in a ferromagnetic-like state at $T_{\rm m}=1.6$ and 0.8 K, respectively.~\cite{Ghimire16}. The magnetic structure of the ordered phase in these compounds remains to be understood. Within the ordered state of Ce$_{2}$CoAl$_{7}$Ge$_{4}$, the magnetization tends to saturate in higher magnetic fields, reaching about 0.55 $\mu_B$ at 7 T for $T=0.6$ K. This lower value of magnetization compared to 2.14 $\mu_B$ for free Ce$^{3+}$ ion is attributed to Kondo/CEF effects. The magnetic entropy at the antiferromagnetic (AFM) transition temperature is approximately 0.27 of $Rln2$, suggesting a strong Kondo effect in the compound. In addition, nuclear magnetic resonance (NMR) studies reported that the Knight shift deviates from the bulk susceptibility below 15 K, confirming the coherent heavy-fermion state at low temperatures~\cite{Dioguardi17}. Therefore, Ce$_{2}$CoAl$_{7}$Ge$_{4}$ is an excellent candidate for high-pressure studies, as the application of pressure is expected to enhance the Kondo effect and drive the system towards a quantum critical point.

Here, we investigate the magnetic structure in the ordered phase of Ce$_{2}$CoAl$_{7}$Ge$_{4}$ using neutron diffraction experiments. Our results reveal a coplanar antiferromagnetic structure with $k = (1,1,1)$. Further, we investigated the evolution of the AFM transition under pressure using electrical transport and ac calorimetry. We observe a suppression of the AFM transition toward zero temperature around 1.1 GPa. Finally, the effect of external pressure on the ground state of Ce$_{2}$CoAl$_{7}$Ge$_{4}$ is compared to that of chemical substitution in the Ce$_{2}$$M$Al$_{7}$Ge$_{4}$ series.

\section{EXPERIMENTAL DETAILS}

Single-crystalline samples of Ce$_{2}$CoAl$_{7}$Ge$_{4}$ were grown using the flux method described in Ref.~\cite{Ghimire16}. The heat capacity at ambient pressure was measured in a Quantum Design Physical Property Measurement System (PPMS) using a quasi-adiabatic thermal relaxation technique. Electrical resistivity measurements were performed on polished single crystals in a standard 4-terminal method in which electrical contacts on the samples were made using 12.5 $\mu$m platinum wires and silver paste. Resistivity was measured using an AC resistance bridge (model 372, Lake Shore) at a measuring frequency of 13.7 Hz. The current was applied in the ab plane. 

Electrical resistivity and heat capacity measurements under hydrostatic pressure were performed using a piston-cylinder type pressure cell with Daphne 7373 oil as the pressure medium. The pressure inside the sample space was determined at low
temperatures using a Pb manometer by measuring the shift in the superconducting transition
temperature. Resistivity was measured using the 4-terminal method employing an ac resistance bridge (model 372, Lake Shore). Resistivity measurements in the temperature range $2-300$~K were performed using a PPMS. The heat capacity was measured on the same sample, but in a separate set-up, using ac calorimetry technique~\cite{Sullivan68}. A constantan wire heater and a chromel-Au/Fe(0.07)\%) thermocouple were attached to the sample using a thin layer of GE varnish. The thermocouple ac voltage was amplified using an SR554 low-noise preamplifier transformer and recorded using an SR860 lock-in amplifier. Measurements were performed at temperatures down to 85 mK in an adiabatic demagnetization refrigerator (Cambridge Magnetic Refrigeration).

Neutron powder diffraction experiments on Ce$_2$CoAl$_7$Ge$_4$ were performed using the DMC diffractometer at the Paul Scherrer Institute, Villigen, Switzerland~\cite{DMC_1990}. About 4~g of loose powdered sample from large single crystals was placed in a He-filled copper can mounted in a dilution refrigerator and measured with $\lambda = 2.453$~\AA\> neutrons at temperatures between $T=0.1$ and $T=3$~K.

\section{Results and Discussion}

\subsection{Magnetic transition}

\begin{figure}[t]
\centering
\includegraphics[width=1\linewidth]{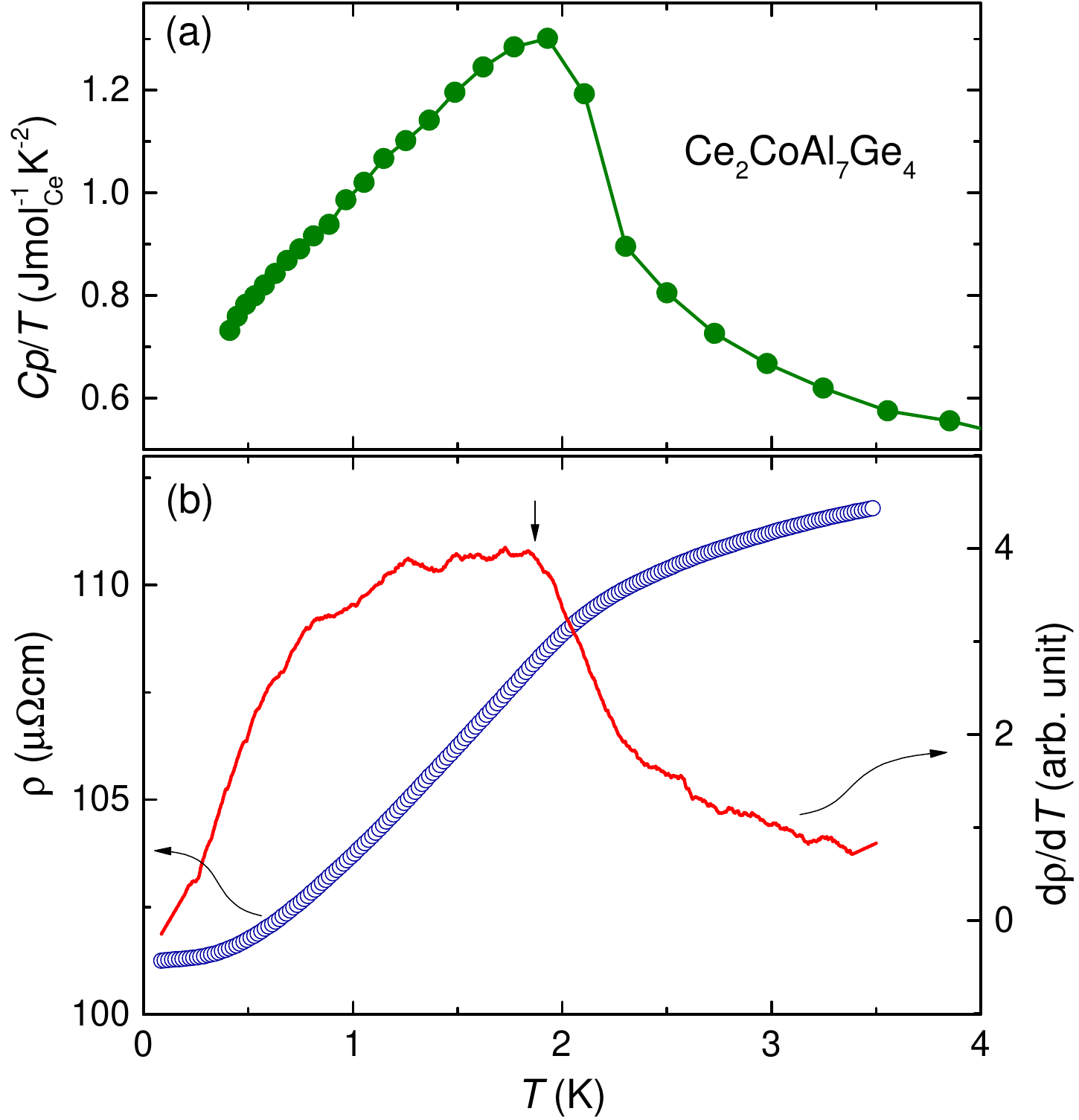}
\caption{(a) Temperature dependence of the heat capacity of Ce$_{2}$CoAl$_{7}$Ge$_{4}$ plotted as $C_p/T$ vs $T$. (b) Low-temperature electrical resistivity, $\rho(T)$, of Ce$_{2}$CoAl$_{7}$Ge$_{4}$ as a function of temperature. Resistivity is measured in a small applied magnetic field of 500~Oe to avoid spurious contributions from superconducting inclusions in the sample. The temperature derivative of resistivity d$\rho(T)$/d$T$ is plotted on the right axis. The downward arrow indicates the transition temperature $T_N$.}
\label{Fig1}
\end{figure}

\begin{figure*}[t]
	\centering\includegraphics[width=1\textwidth]{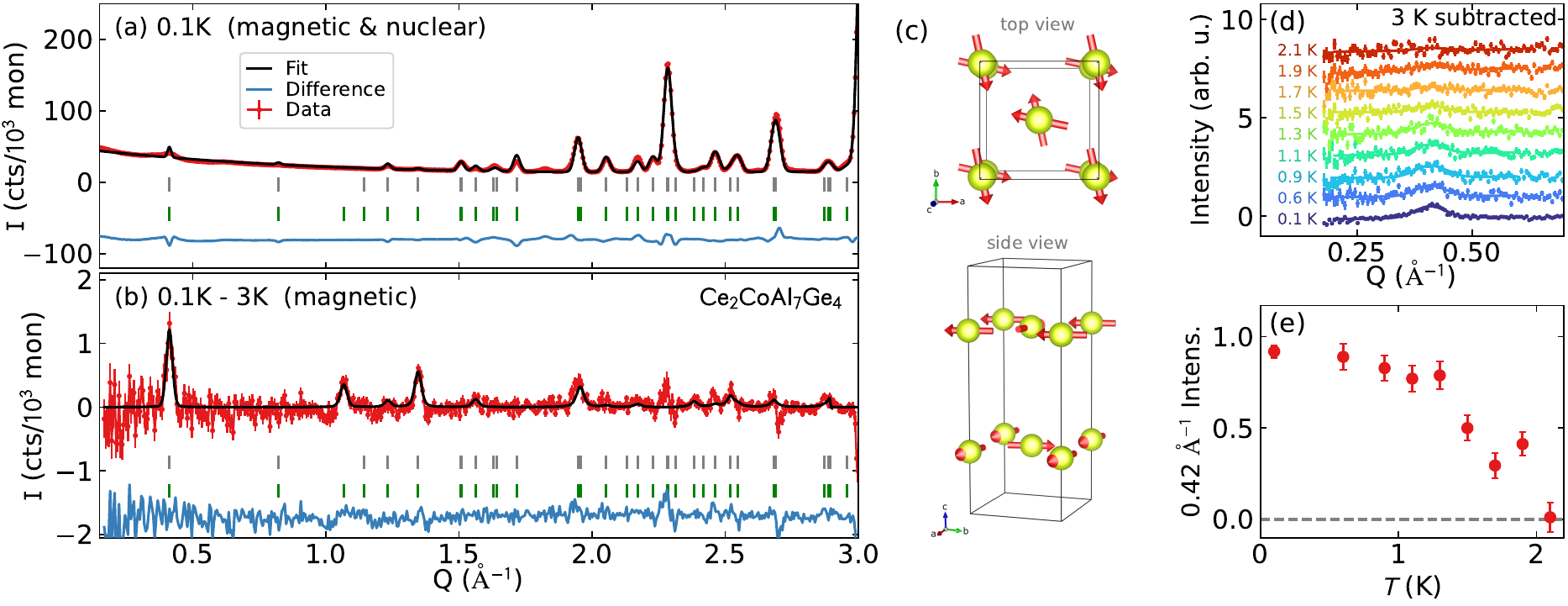}

    \caption{(a) Neutron powder diffraction spectra of Ce$_2$CoAl$_7$Ge$_4$ obtained at $T=0.1$~K. The black solid line is a nuclear refinement fit. (b) The magnetic intensity obtained by subtracting the 3~K pattern from the 0.1~K pattern. The solid black line is the best fit from Rietveld refinement (note that the intensity is two orders of magnitude less than the nuclear refinement). (c) Illustration of the coplanar magnetic structure of the antiferromagnetic phase based on the best fit to the diffraction spectra. (d) The temperature evolution of the 0.42~\AA$^{-1}$ magnetic peak (spectra are offset vertically for clarity). (e) Gaussian fitted intensity of the 0.42~\AA$^{-1}$ peak as a function of temperature.} 
	\label{Fig2}
\end{figure*}

According to previous report, Ce$_{2}$CoAl$_{7}$Ge$_{4}$ undergoes an antiferromagnetic ordering at $T_N=1.8$~K~\cite{Ghimire16}. Figure 1(a) shows the low-temperature heat capacity $C/T$ of Ce$_{2}$CoAl$_{7}$Ge$_{4}$. The peak in $C/T$ corresponds to the AFM transition. Note that the magnetic entropy at $T_N$ is only about 0.27$Rln2$ due to the strong Kondo effect in the compound~\cite{Ghimire16}. The low-temperature part of the electrical resistivity, $\rho(T)$, is plotted on the left axis of Fig.~\ref{Fig1}(b). We note that the flux-grown samples contain a small amount of impurity phases that undergo superconducting transitions at low temperatures. Therefore, resistivity is measured in an applied magnetic field of 500~Oe to avoid the spurious features in resistivity that come from superconducting impurities. The lack of any associated feature in the heat capacity and previous susceptibility measurements confirms that the volume fraction of the impurity phase is rather small. The AFM transition manifests as a shoulder-like anomaly in resistivity owing to the reduction in spin-disorder scattering in the ordered phase. This anomaly is better visible in the temperature derivative of resistivity ($d\rho/dT$) plotted on the right axis of Fig.~\ref{Fig1}(b), where the onset of magnetic transition appears as a sudden increase in $d\rho/dT$. Here, the transition temperature is taken as the temperature corresponding to the maximum in $d\rho/dT$ which also corresponds to the maximum in the heat capacity. We obtain $T_N=1.9$~K, which is consistent with the previously reported value~\cite{Ghimire16}.

\subsection{Magnetic structure}\label{magstr}

In order to investigate the magnetic structure in the antiferromagnetic phase of Ce$_{2}$CoAl$_{7}$Ge$_{4}$, we performed powder neutron diffraction experiments. Figure~\ref{Fig2}(a) shows the diffraction spectra obtained at $T=0.1$~K along with a nuclear refinement fit. The magnetic Bragg peaks are identified by subtracting the spectra obtained at 3~K ($T>T_N$) from the 0.1~K spectra and are plotted in Fig.~\ref{Fig2}(b). At 0.1~K, there are clear changes in intensity relative to the 3~K diffraction pattern, and each of these temperature-dependent Bragg intensity is commensurate with the unit cell. The temperature evolution of the prominent magnetic peak at 0.42~\AA$^{-1}$ presented in Fig.~\ref{Fig2}(d) shows that this temperature-dependent scattering onsets below $T=2$~K. This is clearly seen in the Gaussian-fitted intensity plotted as a function of temperature in Fig.~\ref{Fig2}(e). This temperature-dependent scattering that occurs below $T=2$~K  coincides with the magnetic transition temperature obtained from thermodynamic and electrical transport measurements. We therefore associate this change in Bragg intensity with the magnetic ordering. 

The strongest magnetic Bragg intensity is the $q=(0,0,1)$ peak at 0.42~\AA$^{-1}$. Because magnetic neutron scattering is only sensitive to moments perpendicular to the scattering vector \cite{boothroyd2020principles}, this strongly constrains the magnetic order to be within the $ab$ plane. We used representational analysis and Rietveld refinement (as implemented in the \textit{Fullprof} suite \cite{Fullprof}) to fit the magnetic structure. The best fit to the data comes from an antiferromagnetic order with a wave vector $k=(1,1,1)$.  %(not $k=[0,0,0]$ order, which would be a uniform ferromagnet) 
Rietveld refinements reveal a coplanar magnetic structure shown in Fig. \ref{Fig2}(c) with an ordered moment of $0.383 \pm 0.018 \> \mu_B$ (see the appendix for details).  Note that the structure in Fig.~\ref{Fig2}(c) has a net magnetization in each plane of Ce atoms, but no net magnetization for the unit cell as a whole. The proposed magnetic structure is consistent with previously reported magnetic susceptibility ($\chi$) data, where a large anisotropy in $\chi$ ($\chi$ perpendicular to the $c$ axis greater than $\chi$ along the $c$ axis) was attributed to easy-plane magnetic moments~\cite{Ghimire16}.

\subsection{Pressure evolution of the magnetic transition}\label{n}

\begin{figure}[tb]
\centering
\includegraphics[width=1\linewidth]{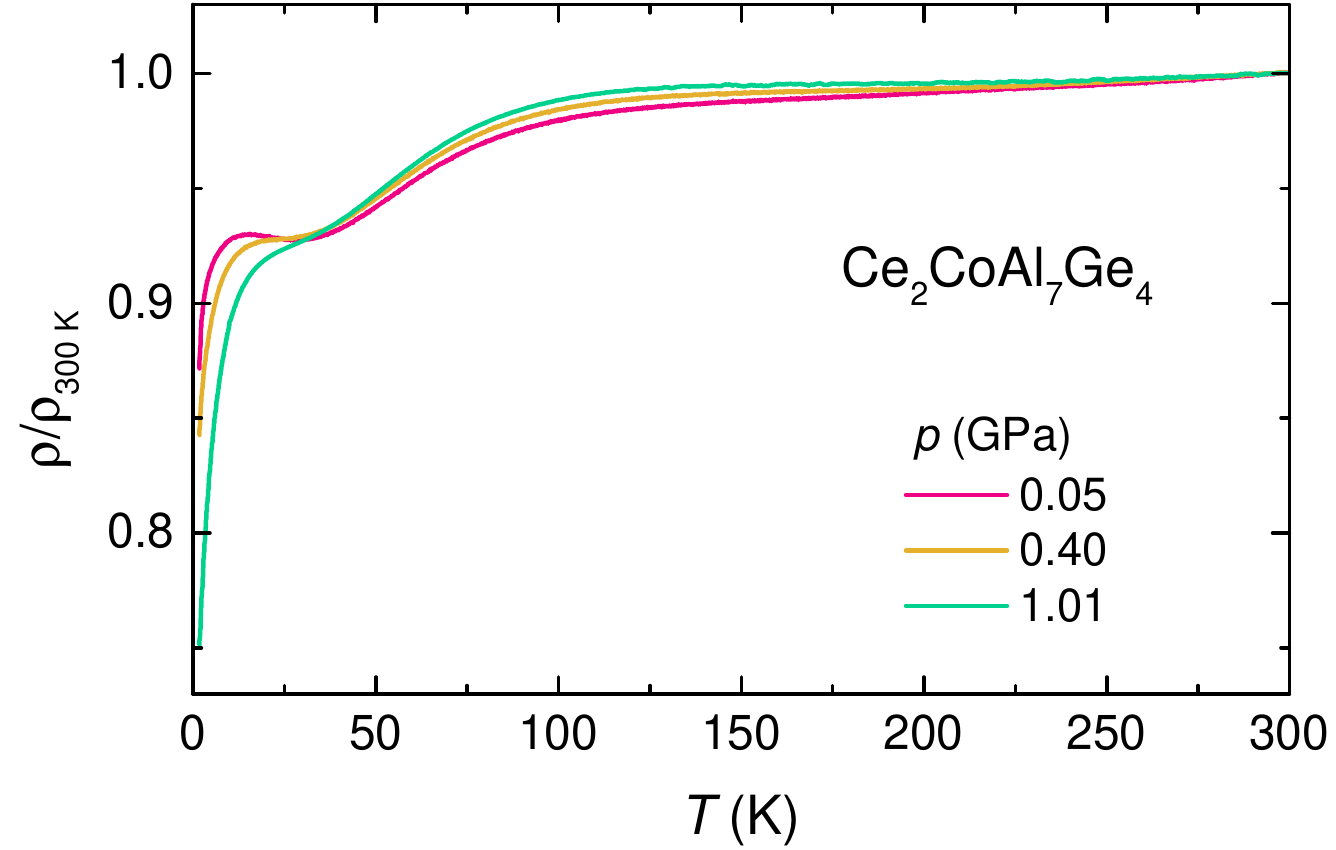}
\caption{Temperature dependence of the normalized resistivity ($\rho/\rho_{\rm 300K}$) of Ce$_{2}$CoAl$_{7}$Ge$_{4}$ for several pressures.}
\label{Fig3}
\end{figure}

The evolution of the magnetic transition in Ce$_{2}$CoAl$_{7}$Ge$_{4}$ under external pressure is investigated using electrical resistivity and ac calorimetry measurements. The temperature dependence of the normalized electrical resistivity, $\rho/\rho_{\rm 300 K}$, for a few selected pressures is plotted in Fig.~\ref{Fig3}. The resistivity shows metallic behavior upon cooling with a broad hump centered around 100~K followed by a minimum around 30~K and a small peak at lower temperature around 15~K. The broad hump around 100~K is likely due to additional scattering originating from thermal population of excited CEF levels. In the case of the isostructural compound Ce$_{2}$PdAl$_{7}$Ge$_{4}$, such broad hump in $\rho(T)$ is observed in a similar temperature range and its excited CEF levels are reported to be at 100 and 156~K, respectively~\cite{Ghimire16}. It is likely that Ce$_{2}$CoAl$_{7}$Ge$_{4}$ has a similar CEF scheme. As shown in Fig.~\ref{Fig3}, the position of the broad hump does not change significantly with increasing pressure. This implies that the CEF splitting is not significantly affected by pressure.   

A small peak in resistivity, centered around 15~K at ambient pressure, is reminiscent of the characteristic peak observed in Kondo systems close to the coherence temperature~\cite{Lassailly85}. Similar features in $\rho(T)$ could also arise from CEF effects~\cite{Cornut72}. However, as mentioned above, the excited CEF levels are likely about 100~K above the ground state. In addition, previous NMR studies have found that the Knight shift deviates from the bulk magnetic susceptibility below $T^*\approx15$~K, providing strong evidence for Kondo coherence below $T^*$~\cite{Ghimire16}. Importantly, the peak in $\rho(T)$ at $T^*$ shows significant pressure dependence and shifts to higher temperatures with increasing pressure. This suggests an enhancement in the Kondo effect in Ce$_{2}$CoAl$_{7}$Ge$_{4}$ under external pressure as expected for Ce-based heavy-fermion systems~\cite{Thompson87, Yomo88}.

\begin{figure}[tb]
\centering
\includegraphics[width=1\linewidth]{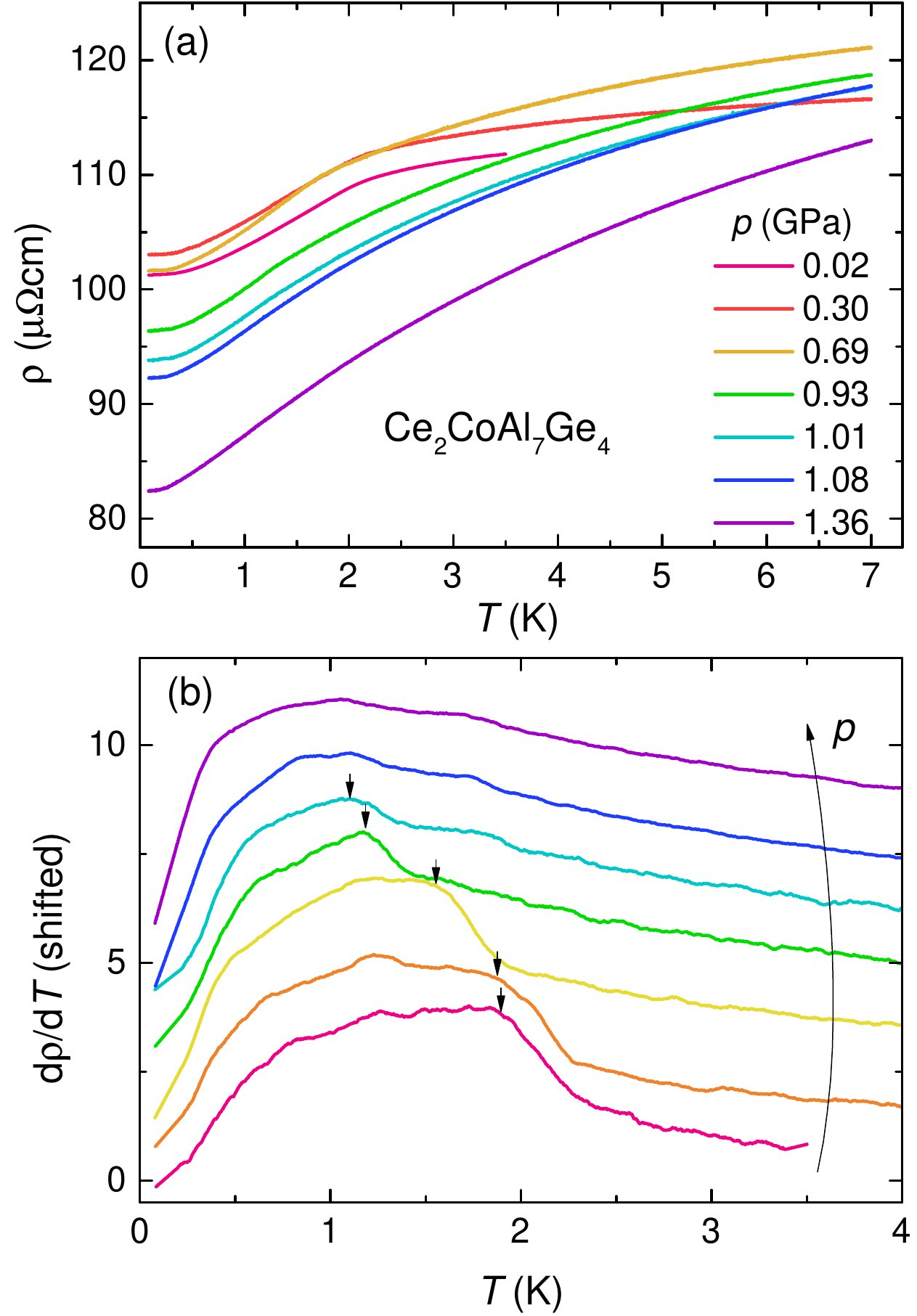}
\caption{(a) Low-temperature $\rho(T)$ \textit{vs.} temperature plot of Ce$_{2}$CoAl$_{7}$Ge$_{4}$ for several applied pressures. (b) Temperature derivative of the resistivity ($d\rho/dT$) as a function of temperature. Note that the curves are shifted along the y-axis for better viewing. The transition temperature $T_N$ is marked by the arrows.}
\label{Fig4}
\end{figure}

Figure~\ref{Fig4}(a) presents the low-temperature resistivity of Ce$_{2}$CoAl$_{7}$Ge$_{4}$ as a function of temperature for several pressures. With increasing pressure, the shoulder-like anomaly in $\rho(T)$ corresponding to the magnetic transition continuously shifts to lower temperatures. As pressure approaches 1 GPa, this anomaly becomes much weaker and eventually disappears at higher pressures. The evolution of this anomaly with pressure is better tracked in the temperature derivative of resistivity ($d\rho/dT$) plotted in Fig.~\ref{Fig4}(b). The magnetic transition appears as a jump in $d\rho/dT$, where the maximum is taken as the transition temperature $T_N$ and is marked by the arrows. $T_N$ is continuously suppressed to lower temperatures with increasing pressure, while the jump in $d\rho/dT$ becomes progressively weaker. $T_N$ reaches about 1 K at $p=1$ GPa. Above this pressure, the feature appears to be washed out, suggesting a quick suppression of the magnetic transition around $p=1.1$~GPa. 

\begin{figure}[t]
\centering
\includegraphics[width=1\linewidth]{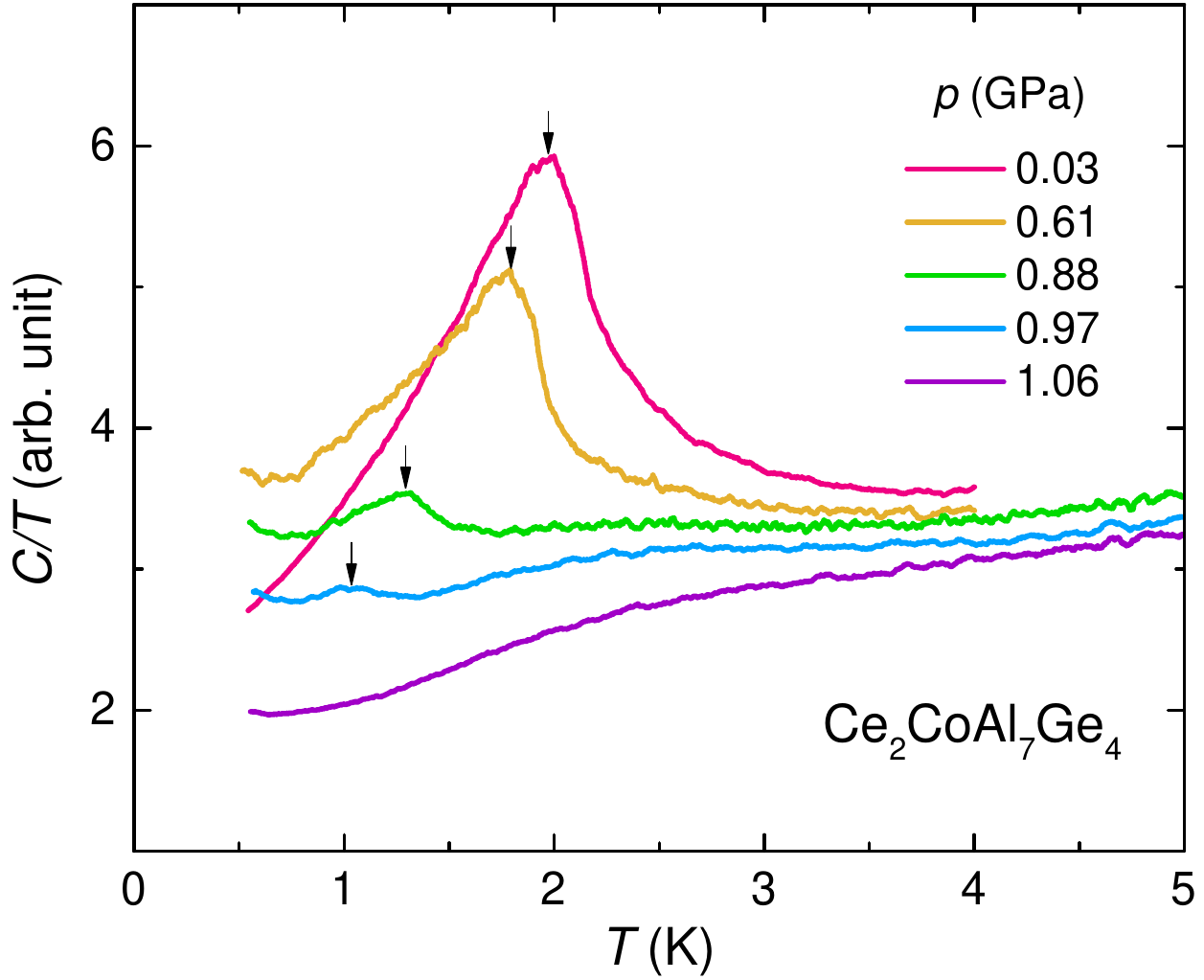}
\caption{Temperature dependence of the heat capacity, obtained via ac calorimetry, of Ce$_{2}$CoAl$_{7}$Ge$_{4}$ for several applied pressures. The curves are shifted vertically for better clarity. The antiferromagnetic transition temperature $T_N$ is marked by the arrows.}
\label{Fig5}
\end{figure}

\begin{figure}[t]
  \centering
  \includegraphics[width=1\linewidth]{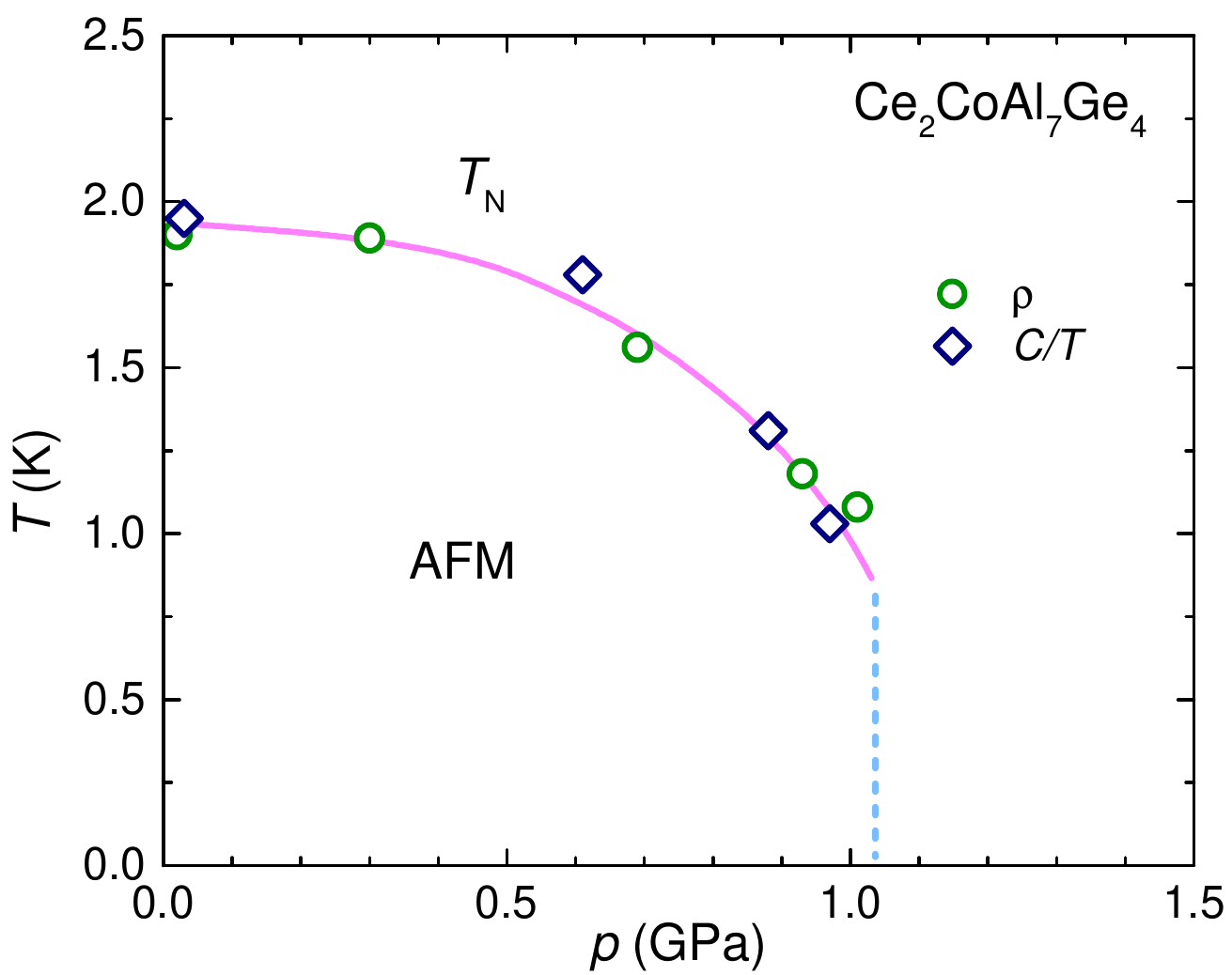}
  \caption{Pressure-temperature phase diagram of Ce$_{2}$CoAl$_{7}$Ge$_{4}$, compiled from electrical resistivity and ac calorimentry data. The lines are guide to the eye. }
  \label{Fig6}
  \end{figure}

To further substantiate the suppression of the antiferromagnetic transition under pressure, we performed ac calorimetry measurements on the same sample. The heat capacity $C/T(T)$ of Ce$_{2}$CoAl$_{7}$Ge$_{4}$ is plotted in Fig.~\ref{Fig5}. The peak in $C/T$ corresponding to the AFM transition shifts to lower temperature with increasing pressure. It is also worth noting that the size of the jump in the heat capacity decreases considerably as the pressure increases. A small peak is visible at $T\approx1$~K for 0.97~GPa, which then completely disappears at higher pressures. All of these observations are consistent with the resistivity data. Combining the resistivity and heat capacity data we construct the temperature-pressure phase diagram of Ce$_{2}$CoAl$_{7}$Ge$_{4}$, presented in Fig.~\ref{Fig6}. The application of pressure suppresses the AFM transition to approximately 1~K at $p\approx1$~GPa. Above 1 GPa, no evidence for a transition is observed, which indicates an abrupt disappearance of the AFM transition. 

A key question to address here is whether the application of pressure suppresses the magnetic transition towards a quantum critical point in Ce$_{2}$CoAl$_{7}$Ge$_{4}$, similar to the effect of chemical substitution of Pd. In Ce$_{2}$PdAl$_{7}$Ge$_{4}$, the proximity to a QCP is suggested based on the absence of magnetic ordering down to 0.4~K and the non-Fermi liquid behavior observed at low temperatures~\cite{Ghimire16}. Signatures of NFL behavior are observed in several physical properties such as $C_{4f}(T)/T\sim -{\rm ln}(T)$, $\rho(T)\sim T$, and $\chi(T)\sim -{\rm ln}(T)$ at low temperatures. Moreover, the application of magnetic field appeared to drive the system towards a Fermi liquid state. In the case of Ce$_{2}$CoAl$_{7}$Ge$_{4}$, close to the apparent critical pressure, the low-temperature resistivity data do not reveal evidence of a systematic change in the temperature dependence of the resistivity from a quadratic to linear behavior. In addition, the heat capacity does not show any logarithmic divergence corresponding to the NFL behavior within the temperature range of the measurements. This is unexpected given that the peak in heat capacity at the magnetic transition rapidly disappears while nearing the critical pressure, releasing most of the magnetic entropy. However, we note that the ac calorimetry technique employed here does not allow for a reliable quantitative comparison of the heat capacity at different pressures. The optimal frequency for the ac calorimetry measurements, which determines the effective thermal coupling between the sample and its surroundings, shows significant temperature and pressure dependence, making a quantitative comparison of the heat capacity data difficult.

Considering the abrupt disappearance of the AFM transition around 1~GPa, a likely scenario is that the AFM transition could become first-order-like and quickly disappear at higher pressures without reaching a QCP. Such a scenario is reported in CeRuPO, where a pressure-induced antiferromagnetic phase shifts to lower temperatures with increasing pressure and then disappears in a first-order-like fashion~\cite{Edit15}.

Another possible scenario is that a QCP is avoided by the onset of a long/short-range magnetic order at higher pressures. The heat capacity data at higher pressures show a broad and weak hump centered around 2-4~K which appears to become more prominent at higher pressures. Such a broad feature in heat capacity could arise from a pressure-induced magnetic phase that is likely developing from short-range correlations. In such a scenario, the missing magnetic entropy from the suppression of the antiferromagnetic order could be redistributed to the pressure-induced phase.  Further investigations, such as NMR and neutron diffraction under pressure, could provide more details and clarify the evolution of magnetism in Ce$_{2}$CoAl$_{7}$Ge$_{4}$.

Upon comparing the application of pressure with the substitution of transition metal, it becomes clear that the chemical substitution (Co, Ir, Ni, and Pd) is not merely a lattice volume effect. While the magnetic ordering temperature consistently decreases, the unit cell volume exhibits a nonmonotonic trend across the series, with values of 537.67, 549.25, 537.86, and 550.76~\AA$^{3}$ for the Co, Ir, Ni, and Pd compound, respectively~\cite{Ghimire16}. In contrast, the lattice parameter $c$ follows a trend similar to that of the magnetic transition temperature; lattice parameter $c$ is similar for Co (15.364~\AA) and Ir (15.366~\AA), and then decreases for Ni (15.273~\AA) and Pd (15.263~\AA). It should also be noted that the lattice parameter $a$ is 1.54\% larger in the Pd compound compared to the Co member. Considering the quasi-2D nature of the material, where Ce square nets are separated by $M$-Al-Ge layers, these anisotropic changes in the lattice parameters probably drive the variation in the magnetic properties across the Ce$_{2}M$Al$_{7}$Ge$_{4}$ series. In contrast, hydrostatic pressure tends to induce more isotropic lattice changes, which may explain why the magnetic behavior evolves differently under pressure compared to Pd substitution.

\section{Summary}
Ce$_{2}$CoAl$_{7}$Ge$_{4}$ is a heavy-fermion compound and undergoes an antiferromagnetic transition below $T_N=1.9$~K. We investigated the magnetic structure of the ordered phase in Ce$_{2}$CoAl$_{7}$Ge$_{4}$ using inelastic neutron scattering. The best fit to the experimental data corresponds to an antiferromagnetic $k = (1, 1, 1)$ order. Rietveld refinements reveal a coplanar magnetic structure, where the Ce moments lie in the $ab$ plane, with an ordered moment of $0.383 \pm 0.018 \> \mu_B$. Additionally, we studied the effect of external pressure on the magnetic transition using electrical transport and ac calorimetry measurements under hydrostatic pressure. Our results show that the antiferromagnetic ordering temperature is suppressed to approximately 1~K at $p\approx1$ GPa, above which the transition abruptly disappears in a first-order-like fashion. We compare the effect of external pressure with the transition metal substitution study on the Ce$_{2}M$Al$_{7}$Ge$_{4}$ series ($M=$Co, Ir, Ni, and Pd), which reported absence of magnetic ordering and proximity to a quantum critical point in Ce$_{2}$PdAl$_{7}$Ge$_{4}$. Our findings highlight that transition metal substitution is not merely a lattice volume effect akin to applied pressure. Our combined results indicate that uniaxial pressure may be a more efficient tuning parameter to drive Ce$_{2}M$Al$_{7}$Ge$_{4}$ compounds towards a quantum critical point.

\section*{Acknowledgements}
 
We thank J.~D.~Thompson for stimulating discussions. We gratefully acknowledge the U.S. Department of Energy, Office of Basic Energy Sciences, Division of Materials Science and Engineering under project ``Quantum Fluctuations in Narrow-Band Systems.'' M.~O.~Ajeesh and Yu Liu acknowledge funding from the Laboratory Directed Research \& Development Program. Neutron diffraction experiments were performed at the Swiss spallation neutron source SINQ, Paul Scherrer Institute, Villigen, Switzerland. Data analysis was supported by the Laboratory Directed Research \& Development Program. 

\section*{appendix: Refinement details}
%\subsection{Refinement details \label{app:Refinement}}
\begin{table*}[tb]
    \centering
    \begin{ruledtabular}
    \begin{tabular}{l|c|c|c|c|cccc}
         Irrep & $\Gamma_1$ & $\Gamma_2$ & $\Gamma_3$   & $\Gamma_4$  & $\Gamma_5$ \\
         Basis vec. & $\psi_1$ &  $\psi_1$ & $\psi_1$ & $\psi_1$   & $\psi_1$ & $\psi_2$  & $\psi_3$ & $\psi_4$  \\ \hline
         site 1  (0,0,-0.26) & $(0,0,1)$  & $(0,0,1)$  & $(0,0,1)$  & $(0,0,1)$     & $(1,0,0)$  & $(0,0,0)$  & $(0,0,0)$  & $(1,0,0)$ \\
         site 2  (0.5,0.5,-0.26) & $(0,0,-1)$ & $(0,0,-1)$ & $(0,0,1)$  & $(0,0,1)$     & $(-1,0,0)$ & $(0,0,0)$  & $(0,0,0)$  & $(1,0,0)$ \\
         site 3  (0,0,0.26) & $(0,0,-1)$ & $(0,0,1)$  & $(0,0,-1)$ & $(0,0,1)$     & $(0,0,0)$  & $(0,1,0)$  & $(0,-1,0)$ & $(0,0,0)$ \\
         site 4  (0.5,0.5,0.26) & $(0,0,1)$  & $(0,0,-1)$ & $(0,0,-1)$ & $(0,0,1)$     & $(0,0,0)$  & $(0,1,0)$  & $(0,1,0)$  & $(0,0,0)$ \\
    \end{tabular}
    \end{ruledtabular}
    \caption{Basis vectors and irreducible representations for magnetic Ce order with wave vector $k=(1,1,1)$. The first four irreducible representations $\Gamma_1$-$\Gamma_4$ have $c$-axis order, whereas $\Gamma_5$ has in-plane order. Note that $\Gamma_5$ actually has eight basis vectors, the four listed above and four more that are identical but have the $a$ and $b$ components swapped.}
    \label{tab:basisvectors}
\end{table*}

\begin{figure}[tb]
	\centering\includegraphics[width=0.5\textwidth]{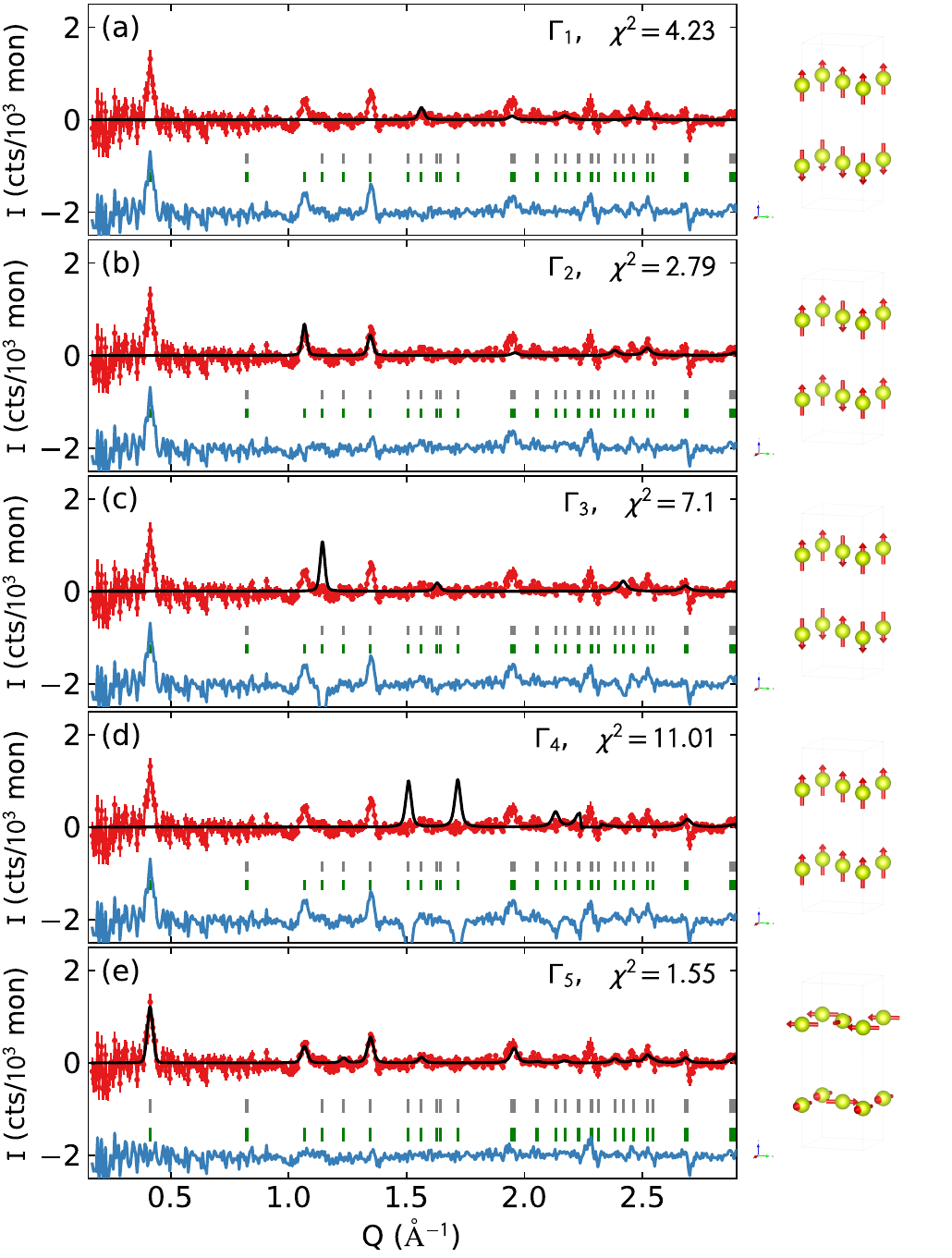}
	\caption{Refinements of Ce$_2$CoAl$_7$Ge$_4$ diffraction data using the five different irreducible representations. The reduced $\chi^2$ is listed in the top right of each panel. The best fit (and the only irreducible representation that captures the 0.42~\AA$^{-1}$ peak) is from $\Gamma_5$ with in-plane order, presented in panel (e).} 
	\label{Fig7}
\end{figure}
The list of basis vectors from irreducible representation analysis is listed in Table \ref{tab:basisvectors}. The refinements from these basis vectors are plotted in Fig. \ref{Fig7}. The only irreducible representation that captures the $q=(0,0,1)$ peak at 0.42~\AA$^{-1}$ is $\Gamma_5$. 
However, because there are eight basis vectors in the coplanar $\Gamma_5$ irreducible representation, the fit is somewhat underdetermined. It is possible to obtain a good fit to the temperature-subtracted data by assuming that the Ce$^{3+}$ sites have two different moment sizes with orthogonal moments (e.g., combining $\psi_1$ and $\psi_3$ with different weights). However, if we assume the Ce sites have equal moment sizes (as one would expect for sites with identical nuclear space group symmetry), the best fit is obtained by combining $\psi_1$ and $\psi_3$ with their $a$-$b$ flipped counterparts, giving the in-plane magnetic structure shown in Fig.~\ref{Fig7}(c). 
The in-plane angle between the spins is governed by the relative intensity of the 0.42~\AA$^{-1}$ peak (the intensity vanishes as the in-plane angle goes to zero), and fits to $(65 \pm 4)^{\circ}$. 

We emphasize that this structure is presented as the most reasonable refinement assuming all spins have equal moments, though it is technically not the only possible refinement. Despite this ambiguity, it is unambiguous that (i) the magnetic ordered structure is $k=(1,1,1)$, (ii) the static moments are fixed to be in-plane, and (iii) the total refined root-mean-squared moment is $\approx 0.38 \> \mu_B$ because this is determined by the total magnetic intensity.

\pagebreak

\bibliography{Ce2174}

\end{document}